\documentclass[sigconf]{acmart}

\AtBeginDocument{%
  }

\setcopyright{acmlicensed}
\copyrightyear{2025}
\acmYear{2025}
\acmDOI{XXXXXXX.XXXXXXX}

\acmConference[Under review]{Under review, 2024}

\usepackage{pgfplots}
\pgfplotsset{compat=newest,
    /pgfplots/ybar legend/.style={
    /pgfplots/legend image code/.code={%
       \draw[##1,/tikz/.cd,yshift=-0.25em, draw=none]
        (0cm,0cm) rectangle (10pt,10pt) ;},
   },
   tick label style = {font = {\fontsize{10 pt}{10 pt}\selectfont}},
    label style = {font = {\fontsize{10 pt}{10 pt}\selectfont}},
    legend style = {font = {\fontsize{10pt}{12 pt}\selectfont}}, 
 }
\usetikzlibrary{patterns,positioning}
\pgfplotstableread[col sep=comma, header=true]{
Model,relevant,somewhat_relevant,not_relevant
M1,60.41,29.29,10.30
M2,77.67,17.34,4.99
M3,72.33,21.40,6.28
M4,72.11,18.74,9.15
M5,81.03,16.52,2.46
M6,66.20,22.07,11.73
M7,72.03,22.61,5.36
 }{\datatableone}

\definecolor{bluegray}{rgb}{0.4, 0.6, 0.8}
\definecolor{asparagus}{rgb}{0.53, 0.66, 0.42}
\definecolor{decentgrey}{RGB}{232,232,232}
\definecolor{awesome}{rgb}{1.0, 0.13, 0.32}
\definecolor{mangotango}{rgb}{1.0, 0.51, 0.26}
\definecolor{persianpink}{rgb}{0.97, 0.5, 0.75}
\definecolor{raspberry}{rgb}{0.89, 0.04, 0.36}
\definecolor{raspberrypink}{rgb}{0.89, 0.31, 0.61}
\definecolor{oceanboatblue}{rgb}{0.0, 0.47, 0.75}
\definecolor{goldenpoppy}{rgb}{0.99, 0.76, 0.0}
\definecolor{brickred}{rgb}{0.8, 0.25, 0.33}
\definecolor{amethyst}{rgb}{0.6, 0.4, 0.8}
\definecolor{deepteal}{RGB}{0, 128, 128}
\definecolor{relevant}{rgb}{0.0, 0.47, 0.75} 
\definecolor{somewhat_relevant}{rgb}{0.99, 0.76, 0.0} 
\definecolor{not_relevant}{rgb}{0.8, 0.25, 0.33} 
	
\usepackage{color, colortbl}
\usepackage{amsmath}

\usepackage[np]{numprint}
\usepackage{booktabs}
\usepackage{siunitx}
\usepackage{multirow}
\usepackage[most]{tcolorbox}
\usepackage{url}
\usepackage{xcolor}
\usepackage{pgfplots}
\usepackage{pgfplotstable}
\usepackage{filecontents}
\usepackage{subcaption}
\usepackage[mode=buildnew]{standalone}
\usepackage{tikz}
\usepackage[eulergreek]{sansmath}
\usepackage{pgfplots}

\definecolor{darkgray}{HTML}{4E4E4E}
\definecolor{gsmblue}{HTML}{3C9CC3}
\definecolor{gsmpink}{HTML}{D94D59}
\definecolor{gsmorange}{HTML}{D97A36}

\usepackage{arydshln}
\tikzset{SubCaption/.style={
text width=2in,yshift=-3mm, align=center,anchor=north
}}

\usepgfplotslibrary{colorbrewer, groupplots}
\usepackage{xspace}

\newcommand{\fqt}[1]{``#1''}
\newcommand{\fqtperiod}[1]{``#1.''\xspace}

\begin{document}

\title{Aug2Search: Enhancing Facebook Marketplace Search with LLM-Generated Synthetic Data Augmentation}

\author{Ruijie Xi}
\affiliation{%
  \institution{North Carolina State University}
  \city{Raleigh}
  \state{North Carolina}
  \country{USA}
}
\email{rxi@ncsu.edu}

\author{He Ba}
\affiliation{%
  \institution{Meta}
  \city{Bellevue}
  \state{Washington}
  \country{USA}
}
\email{bach@meta.com}

\author{Hao Yuan}
\affiliation{%
  \institution{Meta}
  \city{Bellevue}
  \state{Washington}
  \country{USA}
}
\email{hayuan@meta.com}

\author{Rishu Agrawal}
\affiliation{%
  \institution{Meta}
  \city{Bellevue}
  \state{Washington}
  \country{USA}
}
\email{rishu@meta.com}

\author{Yuxin Tian}
\affiliation{%
  \institution{Meta}
  \city{Bellevue}
  \state{Washington}
  \country{USA}
}
\email{yuxin1211@meta.com}

\author{Ruoyan Kong}
\affiliation{%
  \institution{Meta}
  \city{Bellevue}
  \state{Washington}
  \country{USA}
}
\email{ruoyankong@meta.com}

\author{Arul Prakash}
\affiliation{%
  \institution{Meta}
  \city{Bellevue}
  \state{Washington}
  \country{USA}
}
\email{arulprakash@meta.com}


\begin{abstract}

Embedding-Based Retrieval (EBR) is an important technique in modern search engines, enabling semantic match between search queries and relevant results. However, search logging data on platforms like Facebook Marketplace lacks the diversity and details needed for effective EBR model training, limiting the models' ability to capture nuanced search patterns. To address this challenge, we propose Aug2Search, an EBR-based framework leveraging synthetic data generated by Generative AI (GenAI) models, in a multimodal and multitask approach to optimize query-product relevance.

This paper investigates the capabilities of GenAI, particularly Large Language Models (LLMs), in generating high-quality synthetic data, and analyzing its impact on enhancing EBR models. We conducted experiments using eight Llama models and 100 million data points from Facebook Marketplace logs. Our synthetic data generation follows three strategies: (1) generate queries, (2) enhance product listings, and (3) generate queries from enhanced listings. 

We train EBR models on three different datasets: sampled engagement data or original data ((e.g., \fqt{Click} and \fqt{Listing Interactions})), synthetic data, and a mixture of both engagement and synthetic data to assess their performance across various training sets.
Our findings underscore the robustness of Llama models in producing synthetic queries and listings with high coherence, relevance, and diversity, while maintaining low levels of hallucination.
Aug2Search achieves an improvement of up to 4\% in ROC\_AUC with 100 million synthetic data samples, demonstrating the effectiveness of our approach. 
Moreover, our experiments reveal that with the same volume of training data, models trained exclusively on synthetic data often outperform those trained on original data only or a mixture of original and synthetic data.

\end{abstract}

\keywords{Generative Artificial Intelligence, Embedding-based Retrieval, E-commerce, Large Language Models}

\maketitle

\section{Introduction}

Facebook Marketplace is a global e-commerce platform that facilitates interactions between buyers and sellers worldwide. 
With hundreds of millions of products available, the platform supports a vast and diverse array of listings, encompassing everything from furniture and electronics to automobiles and rental properties. 
This scale, along with the variety of products, poses unique challenges for the search system, which must efficiently match user queries with relevant listings to create a seamless shopping experience. 
As such, the Marketplace search engine requires advanced retrieval capabilities to capture the complexity and diversity of user queries, linguistic nuances, and product details.

\begin{figure}
    \centering    
    \includegraphics[clip, trim=1cm 7.5cm 11cm 5.5cm, scale=0.5]{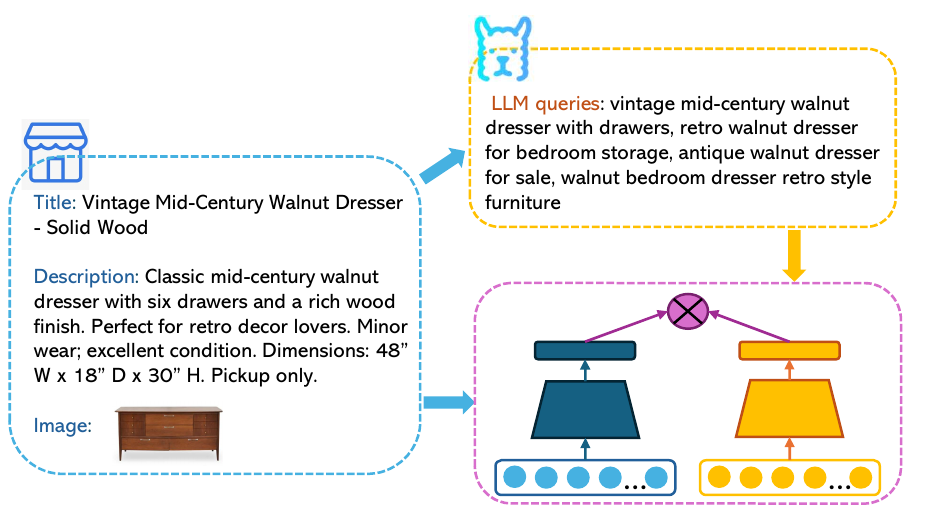}
    \caption{Overview of Aug2Search: For each Facebook Marketplace listing, N synthetic search queries are generated using a GenAI model (Llama). They are both fed to the two-tower SparseNN models for representation learning. In addition, we also enhance product listings with GenAI models. 
    }
    \label{fig:teaser}
    \Description{}
\end{figure}

Facebook Marketplace has adopted embedding-based retrieval (EBR) models to enhance search by bridging the semantic gap between user queries and product listings. EBR represents queries and listings as dense embeddings in a shared space, enabling accurate, context-sensitive retrieval despite differences in vocabulary and phrasing~\cite{huang-2020-embedding}. Advances like Que2Search have improved query and document understanding efficiency~\cite{liu-2021-que2search}, while Que2Engage focuses on surfacing relevant and engaging products~\cite{he-2023-que2engage}. MSURU leverages weakly supervised data for large-scale image classification~\cite{tang-2019-msuru}, and HierCat provides hierarchical query categorization to handle Marketplace's diverse queries with weak supervision~\cite{he-2023-hiercat}. 
These models collectively strengthen EBR in Marketplace. However, challenges persist around data quality variability and the need to adapt to evolving user behavior.

Traditional training data lacks the diversity and detail needed to capture the nuances in user search queries and product listings, limiting the models' ability to perform accurate semantic matching.
For example, product listings provided by sellers often lack standardization, contain typographical errors, and vary widely in style and detail. 
Moreover, reliance on historical data introduces potential biases that can prevent the model from effectively adapting to emerging product categories and capturing shifts in user search behavior.

To address the challenges of data variability, we introduce a flexible and adaptive framework, named Aug2Search, to enhance EBR models through synthetic data augmentation. 
Figure~\ref{fig:teaser} outlines the Aug2Search framework- which leverages large, pre-trained GenAI models to generate synthetic queries and enhance product listings, capturing the diversity and complexity of real-world search interactions. 
Aug2Search incorporates a two-tower, multimodal, multistage, and multitask approach to optimize query relevance and search performance, adapting an XLM encoder~\cite{conneau-2019-cross} to process the query and the text from the listing. 
The model integrates a shared Multi-Layer Perceptron (MLP) and deep sets fusion to create an image modality token by adapting a transformer-fusion mode\cite{yu-2022-commerce}. 

To evaluate Aug2Search, we first assess the quality of the GenAI-generated synthetic data through standard metrics and human evaluations. We experiment with and compare eight GenAI models to assess their ability to produce high-quality synthetic data. Next, we evaluate the performance of the EBR model by training on original data, synthetic data, and a combination of both. 

Our findings confirm the robustness of Meta's GenAI models, particularly Llama3 and Llama3.1, in generating synthetic queries and listings with high coherence, relevance, and diversity, while maintaining low levels of hallucination. Results indicate that Aug2Search improves ROC\_AUC by up to 4\% with 100 million synthetic data points. 
In addition to this, our experiments show that increasing the volume of training data consistently enhances EBR performance, with models trained solely on synthetic data often outperforming those trained on a mix of original and synthetic data.

\begin{figure*}
    \centering
    \includegraphics[clip, trim=0cm 10cm 4.5cm 3cm, scale=0.7]{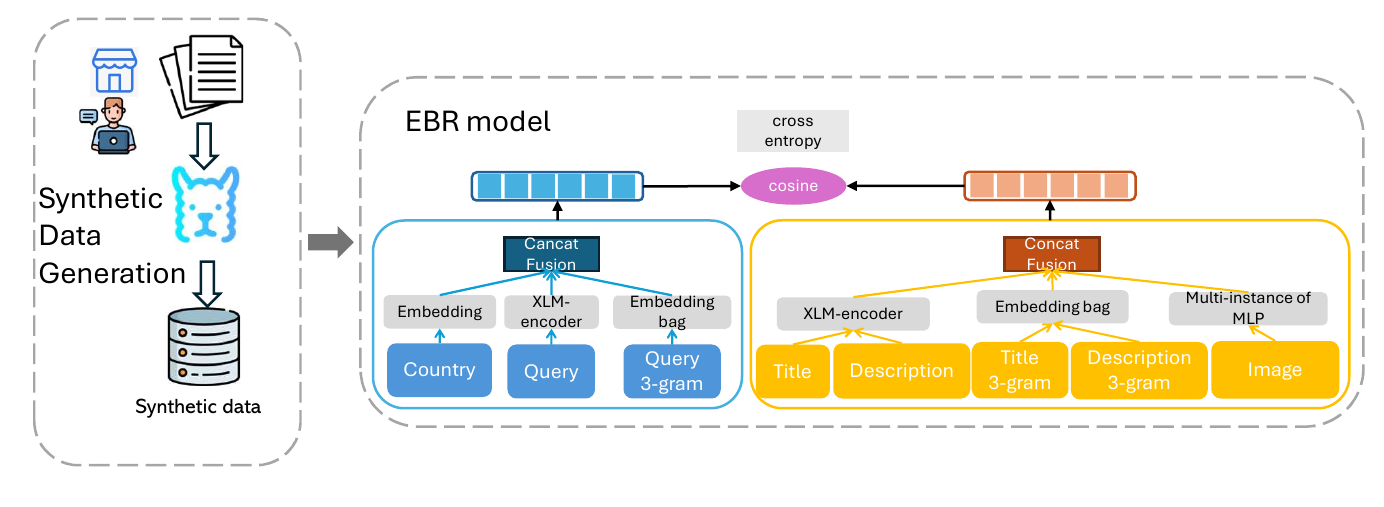}
    \caption{Framework of Aug2search.}
    \label{fig:framework}
    \Description{}
\end{figure*}

\section{Related Work}
\paragraph{Information Retrieval in E-Commerce} Information retrieval (IR) in e-commerce often employs embedding-based methods to align query and item spaces, thereby enhancing retrieval accuracy \cite{li-2023-learning}. Generative retrieval models have demonstrated strong performance on large benchmarks \cite{zeng-2024-scalable}. Multi-level matching networks bridge the gap between informal queries and categories, boosting classification \cite{yuan-2023-multi}. 
Moreover, EBR models have been adopted in e-commerce search to retrieve semantically relevant products as a complement to lexical retrieval \cite{li-2021-embed, magnani-2022-semantic}.
HierCat utilizes multi-task learning to handle intent ambiguity and data imbalance, improving relevance and engagement \cite{he-2023-hiercat}. 
Que2Search and Que2Engage optimize multimodal and multitask query-product representation and balance relevance and engagement \cite{liu-2021-que2search, he-2023-que2engage}. 
Recent advances in reranking and truncation integrate these tasks, achieving state-of-the-art results \cite{xu-2024-list}.

\paragraph{Generating Synthetic Data for E-Commerce Systems} Synthetic data generation enhances e-commerce retrieval and query coverage. Large language models (LLMs) generate synthetic queries to improve responsiveness and capture complex user queries \cite{rome-2024-ask, sannigrahi-2024-synthetic}. 
In e-commerce, LLM-generated queries align better with user intent, as evidenced by increased Click-Through Rates (CTR) in applications such as  Alipay \cite{chen-2024-llmgr}. 
They also expand search capabilities for new categories, bridging semantic gaps in long-tail queries \cite{li-2022-query, jagatap-2024-improving}. Additionally, LLMs support product attribute extraction and increase relevance in listings \cite{fang-2024-llm}. Fine-tuned models in creative tool searches further enhance accuracy and engagement \cite{kumar-2023-multi}. These studies underscore the value of LLM-driven synthetic data for improving query understanding and data diversity in e-commerce \cite{xu-2024-retrieval}.

\section{Methods}
\label{sec:methods}

This section presents Aug2Search (illustrated in Figure~\ref{fig:framework}), a flexible framework that adapts synthetic data augmentation to enhance EBR models in Facebook Marketplace. 
We outline the architecture of the Facebook Marketplace search model and describe our integration of GenAI-generated synthetic data within this framework.

\subsection{Facebook Marketplace Search Framework Overview}

The Facebook Marketplace Search Framework leverages an EBR architecture designed to capture semantic relevance and enhance user engagement \cite{huang-2020-embedding,liu-2021-que2search, he-2023-hiercat,he-2023-que2engage}. 
This framework is implemented using a two-tower model architecture, where the \textit{Query Tower} and \textit{Document Tower} independently encode user queries and product listings into a unified embedding space. 
This section details the components and configurations used in each tower, as well as the approaches applied to optimize search performance.

\subsubsection{Query Tower}

The Query Tower generates multi-granular representations by processing raw query text, character trigrams, and country information. The raw text is encoded via a two-layer XLM model \cite{conneau-2019-cross}, while character trigrams--created by hashing three-character sliding windows--are encoded with an EmbeddingBag layer using sum pooling to capture character-level nuances \cite{niu-2021-dual}. 
An additional EmbeddingBag layer is applied to encode the country feature. 
The final query representation is obtained by fusing these three embeddings through attention-weighted aggregation.

\subsubsection{Document Tower}

The document tower encodes product listings by integrating multiple modalities, including textual data (title and description), images, and contextual information such as price, category, and creation time. 
Contextual features are processed through an MLP-based encoder, where numerical attributes map to individual neurons and categorical attributes are one-hot encoded. 
A BatchNorm layer and a final MLP ensure consistent scaling and dimensionality, producing a "context token" that represents contextual metadata. 
This context token is then incorporated as an additional modality alongside text and image data during the fusion process.
We also apply an EmbeddingBag \cite{niu-2021-dual} layer to encode the character 3-grams multi-categorical features of title and description, as described in the previous paragraph. 
For each document, there are a variable number of images attached to it. 
We take the pre-trained image representations \cite{bell-2020-groknet} for each of the attached images.

\subsubsection{Multimodal Fusion}

The text encoder converts product titles and descriptions into word tokens, embedding them via a transformer with a [CLS] token for sequence representation. 
For attached images, pre-trained representations are enhanced through a shared MLP and deep sets fusion to create an image modality token. 
Our architecture uses a transformer-fusion model \cite{yu-2022-commerce} that merges text, image, and context tokens in a multimodal fusion encoder, initialized from a six-layer XLM-R model \cite{conneau-2019-cross}. 
The text encoder inherits the initial $K$ layers, while the fusion encoder inherits the remaining $M$ layers. 
The final document embedding is obtained from the projected [CLS] token at the last fusion layer, with embeddings fused using attention weights for the final document representation.

\subsection{Synthetic Data Generation}

Instruction-tuned LLMs are primarily accessible through fixed, black-box application programming interfaces (APIs), which restrict direct parameter fine-tuning but enable high-quality text generation via \textit{prompting}. Effective prompt design is crucial: prompts must be concise, task-oriented, and context-rich to align model outputs with specific objectives.
Therefore, we carefully crafted prompts to produce diverse and relevant synthetic data, enhancing the representation of both the queries and the listings.

\subsubsection{Static prompting} 
\label{sec:prompt}
We employ static prompts with fill-in-the-blank templates that are parameterized by specific product characteristics to guide the generation. These templates enable us to generate variations in user search queries and enhance product listings, allowing us to tailor each prompt to capture a wide range data while maintaining consistency in model responses.

\begin{tcolorbox}[colback=darkgray!5!white, colframe=darkgray, label=templates, title=Two Gradually Improved Templates for Query Generation]

\textbf{\textcolor{gsmblue}{T1-Basic:}} Your task is to generate queries for each product a Facebook Marketplace shopper would search for on an e-commerce website. 
The queries should consist of short phrases or keywords and include details of the product, such as brand, color, size, material, condition, and location.

\smallskip
\scalebox{0.9}{\colorbox{amethyst!30}{\{title\}}} \\
\scalebox{0.9}{\colorbox{amethyst!30}{\{description\}}}

\tcbline

\textbf{\textcolor{mangotango}{T2-Detailed:}} Your task is to generate queries for products a Facebook Marketplace shopper would search for on an e-commerce website. 
The queries should consist of short phrases or keywords. Let's think \textcolor{raspberry}{step-by-step}:
\begin{enumerate}
    \item Thoroughly analyze the \#Title and \#Description provided by a Facebook Marketplace seller, correcting any grammar or spelling errors to ensure accurate understanding.
    \item Carefully identify the primary products that accurately represent the title and description, focusing on the most relevant items to avoid confusion.
    \item Craft a diverse array of search queries (\#Queries) pertinent to the products, taking into account different combinations of attributes and alternative ways buyers might express their search intentions (e.g., synonyms, abbreviations).
\end{enumerate}
\smallskip
\scalebox{0.9}{\colorbox{amethyst!30}{\#Title: \{title\}}} \\
\scalebox{0.9}{\colorbox{amethyst!30}{\#Description: \{description\}}}

\end{tcolorbox}

\paragraph{Query generation}
To identify the most effective prompt template, we conducted preliminary experiments evaluating query generation outcomes. 
Based on these results, we iteratively adapted the templates to optimize synthetic data generation. Although multiple intermediate templates were tested, we present only the initial and final versions in this paper.
As shown in \ref{templates}, \textbf{T1-basic} prompts the GenAI model to generate straightforward, attribute-focused queries that mirror typical user search behavior, emphasizing key product details such as brand, model, and features. 
Building on T1 and intermediate templates, \textbf{T2-detailed} guides the model through a \textit{step-by-step} analysis to enhance relevance and diversify queries \cite{kojima-2022-step}. 
Targeted prompts, such as \textit{thoroughly analyze} and \textit{craft a diverse list of queries}, capture a broader range of product attributes, while the \fqt{\#} symbol is used to mark the beginning and end of titles and descriptions for structured input.
We focused on extracting and emphasizing specific product attributes (e.g., color, size, condition) as well as associative categories that users might search for. Associative queries are crafted to include comparable or alternative products that broaden potential search relevance; for example, a listing for \textsl{Nintendo Switch} may generate queries related to \textsl{PS4} or \textsl{Xbox One} consoles.

\paragraph{Listing enhancement}
Building on the strong performance of T2-Detailed, we used this approach to enhance product listings, improving data quality by refining descriptions, correcting errors, and clarifying incomplete information for greater consistency. 
In practice, we found that many listings lack detail or contain irrelevant terms (e.g., "Obo", "Pick up Only", etc.). 
For missing attributes (e.g., vehicle trims), we employed a GenAI model to enrich content and rephrase text for clarity. To mitigate LLM hallucinations \cite{maynez-2020-faithfulness}, we crafted stricter prompts to limit responses to features provided in the input only, avoiding inferred details (e.g., vehicle mileage). 
We also refined prompts to prevent non-factual or promotional language (e.g., "a great opportunity") and avoided inserting placeholder text when descriptions were incomplete. 
The prompt details are provided in Appendix~\ref{app:prompt_listings}.

\section{Experiments}
In this section, we outline the experimental setup used to evaluate the performance of Aug2Search, focusing on training data generation, model training, and evaluation methodologies.

\subsection{Model Training}
\label{sec:training}

Our training data consists of original query-document pairs combined with varying mixtures of synthetic data, which we assess using relevance metrics.  

\subsubsection{Training Data}
\label{sec:strategies}

We evaluate Aug2Search using an internal proprietary dataset of 100 million de-identified, aggregated queries and product listings (titles and descriptions, referred to as \textsl{documents}) and user product engagement signals (e.g., \fqt{Click} and \fqt{Listing Interactions}) from Facebook Marketplace search logs. 
Our pipeline generates synthetic data by first inputting aggregated query-listing pairs into GenAI models, as described in Section~\ref{sec:prompt}, and then blending original and augmented entries to form varied training sets.

\paragraph{Generate synthetic datasets:}
To create robust synthetic data samples for our experiments, we implement three complementary strategies: (1) S1: generating queries directly from original listings, (2) S2: enhancing the listings to enhance content quality and coherence, and (3) S3: enhancing listings and subsequently generating queries based on the enhanced content. 
When generating queries, we input a product listing's title and description from Facebook Marketplace search logs into GenAI models, which produce structured outputs via Meta's internal API. 
We request one to ten queries per listing and observe that the model consistently generates the maximum of ten. 
When enhancing listings, we provide a product listing's title and description and ask the model to enhance the description by correcting errors, adding details, and improving clarity.
If only a title is provided, the model expands it with relevant information, yielding a more complete and informative listing.

\subsubsection{Contrastive Learning}

To optimize for semantic relevance in embedding-based retrieval (EBR), we employ contrastive learning with batch negative sampling, as described in \cite{liu-2021-que2search, he-2023-que2engage}. Positive pairs consist of user-engaged <query, product> interactions, while negatives are sampled from other products within the batch. 
To illustrate how batch negatives work, we describe the process within a single training batch of size \( B \). 
In each batch, we obtain the query embedding tensor \(\{q_i\}_{i=1}^B\) and the document embedding tensor \(\{d_j\}_{j=1}^B\), both with embedding dimension \( D \). 
We then compute a cosine similarity matrix \(\{\cos(q_i, d_j)\}_{i,j=1}^B\), which captures query-document similarity for all pairs in the batch, with rows representing queries and columns representing documents.
This setup frames a multi-class classification task with \( B \) classes, where document \( d_i \) is the correct match for query \( q_i \), and other documents \( d_j \), \( j \neq i \), serve as negative examples. 
We use a scaled multi-class cross-entropy loss to optimize the network, following the scaling approach of Deng et al. \cite{deng-2019-arcface} and setting the scale \( s \) between 15 and 20 as recommended in \cite{liu-2021-que2search}.
The relevance loss function is defined as:
\begin{equation}
    L_{\text{relevance}} = -\frac{1}{B} \sum_{i=1}^{B} \log \left( \frac{\exp \{ s \cdot \kappa (q_i, d_i) \}}{\sum_{j=1}^{B} \exp \{ s \cdot \kappa (q_i, d_j) \}} \right)
\end{equation}
where $B$ is the batch size, $\kappa$ denotes cosine similarity, and $s$ is a scaling factor set to 20. This setup encourages the model to maximize similarity for true query-product pairs while minimizing it for non-relevant products in the batch.

\subsubsection{Learning Contextual Information}

Standard batch negatives are effective for learning semantic relevance but insufficient for capturing engagement-driving contextual factors, such as product price, which play a crucial role in determining user interest. Since batch negatives are sampled from engaged pairs, they lack the diversity needed to represent unengaging yet contextually relevant products.
To address this, we incorporate an auxiliary training objective that includes hard negatives--\texttt{<query, product>} pairs displayed to users but that received no engagement. We define the engagement loss \( L_{\text{engagement}} \) as:
\begin{equation}
    L_{\text{engagement}} = -\left( y_i \log(c_i) + (1 - y_i) \log(1 - c_i) \right)
\end{equation}
where \( y_i \) is the binary engagement label, \( c_i = s \cdot \kappa (q_i, d_i) \), and \( s \) is a scaling factor.

To combine semantic relevance and engagement-driven optimization, we define the final multitask loss as:
\begin{equation}
    L(\theta) = \lambda_1 \cdot L_{\text{relevance}} + \lambda_2 \cdot L_{\text{engagement}}
\end{equation}
where \( \theta \) represents the model parameters, and \( \lambda_1 \) and \( \lambda_2 \) are empirically chosen weighting parameters. This multitask approach ensures that the model captures both relevance and engagement nuances, optimizing for user interaction as well as semantic alignment.

\subsubsection{Training Parameters}
All models are trained on Nvidia A100 GPUs with a batch size of 512, using the Adam optimizer with a learning rate of $2 \times 10^{-4}$. We set warmup steps to 2000, with different learning rates for the XLM encoder ($2 \times 10^{-4}$) and other components ($7 \times 10^{-4}$). Regularization techniques include dropout (rate of 0.1) and gradient clipping (value of 1.0). 
ROC\_AUC on the validation set serves as the early stopping metric with a patience of three epochs. 
Following \cite{he-2023-que2engage}, we set $\lambda_1$ and $\lambda_2$ as 0.2 and 0.8, respectively.

\subsection{Evaluation}
In Facebook Marketplace, we collect human-rated data to evaluate search engine quality. Human evaluation is conducted on publicly visible products, with query data de-identified and aggregated beforehand \cite{liu-2021-que2search, he-2023-que2engage}. 
For our evaluation, we submitted a set of sampled queries with search results for human rating. 
Human raters labeled each retrieved listing as \texttt{Relevant} (2), \texttt{Somewhat Relevant} (1), or \texttt{Off-topic} (0), yielding an evaluation dataset of \np{34000} (query, document) pairs. 
This human-rated data allows us to assess relevance and overall search quality effectively.
Our evaluation focuses on several metrics to measure semantic relevance alignment between query and listing embeddings.

\paragraph{ROC\_AUC}
Following previous works \cite{liu-2021-que2search, he-2023-que2engage}, we use ROC\_AUC to quantify the model's ability to distinguish relevant from irrelevant results, based on human-labeled relevance. 
This metric serves as both an offline evaluation measure and a training validation metric, guiding early stopping. 
Using the cosine similarity between an inferred query-document embedding pair, $\cos(q, d)$, as a score, we calculate ROC\_AUC. 
Evaluating ROC\_AUC on the evaluation dataset provides a robust indicator of the model's alignment with search relevance, reflecting consistency between search retrieval and ranking \cite{he-2023-que2engage}.

\paragraph{Points Biserial Correlation (PBC)}
Points Biserial Correlation (PBC) evaluates the correlation between 
listings and queries embeddings and binary relevance labels \cite{kornbrot-2014-point}. Given that our relevance label includes three distinct categories, we convert the ternary labels into binary labels to facilitate our analysis.
Two schemes, \textsl{PBC-o} and \textsl{PBC-r}, provide nuanced assessments by defining binary labels differently: \textsl{PBC-o} assigns 0 to \texttt{Somewhat Relevant} and \texttt{Off-topic}, while \textsl{PBC-r} assigns 1 to \texttt{Somewhat Relevant} and \texttt{Relevant}. 
Higher PBC values signify stronger alignment with human ratings.

\paragraph{Relevance Consistency Ratio (RCR)}
The Relevance Consistency Ratio (RCR) measures the model's precision in distinguishing relevant from irrelevant listings. 
Calculated as the ratio of irrelevant to relevant pairs with cosine similarity scores exceeding the median \texttt{Off-topic} similarity, a lower RCR implies that high similarity values align closely with genuine relevance. 
This metric offers insight into the model's scoring consistency, supporting enhanced search quality and user satisfaction.

\section{Results}

We first examined the performance of various Meta GenAI models (as detailed in Table~\ref{tab:models}), followed by an evaluation of Facebook Marketplace's EBR models. 
This dual-stage evaluation allows us to assess the effectiveness of GenAI models in enhancing query-to-product retrieval through embedding-based methods. 

\begin{table}[!htb]
    \centering
    \begin{tabular}{ll}
    \toprule
        Model & Description \\
    \midrule
        M1 & Llama2-70b-chat \\
        M2 & Llama3-8b-instruct \\ 
        M3 & Llama3-70b-instruct \\ 
        M4 & Llama3.1-8b-instruct \\ 
        M5 & Llama3.1-70b-instruct \\ 
        M6 & Llama3.2-1b-instruct \\ 
        M7 & Llama3.2-3b-instruct \\
        M8 & Llama3.1-405b-instruct\\
    \bottomrule
    \end{tabular}
    \caption{GenAI models used for synthetic data generation.}
    \label{tab:models}
\end{table}

\subsection{GenAI's Effectiveness in Generating High-Quality Synthetic Data}

We deployed a range of metrics to assess the quality of synthetic data, focusing on the relevance of generated queries and the quality of enhanced listings. 
To ensure a robust evaluation, we avoided lexical-overlap-based metrics such as ROUGE \cite{lin-2004-rouge}, which are insufficient for evaluating the quality of long-form answers generated by LLMs \cite{krishna-2021-hurdles, kamalloo-2023-evaluating}. 
We also excluded contextual similarity measures like BERTScore \cite{zhang-2019-bertscore}, as our goal is not merely to capture semantic similarity but to enrich the model training data.

For query evaluation, we randomly selected 50 product listings from the original training data, generating 500 queries\footnote{Each product listing prompted the GenAI model to generate exactly ten queries, following a Meta-internal schema.}. 
For listing evaluation, we selected 200 distinct listings from the training dataset. 
To assess the alignment quality of GenAI data and human values, we followed the alignment criteria from \citet{askell-2021-general}: an assistant is considered aligned if it is helpful, honest, and harmless (3H):

\begin{enumerate}
    \item \textbf{Helpfulness}: The model effectively assists users by accurately addressing their queries.
    \item \textbf{Honesty}: The model provides truthful information, acknowledges uncertainty where applicable, and avoids misleading responses.
    \item \textbf{Harmlessness}: The model refrains from generating harmful content, including hate speech or violent language.
\end{enumerate}

Based on the 3H criteria and Facebook Marketplace's commerce policies\cite{fb_commerce_policy}, we designed questions to align with our research goals for both human raters and LLMs. 
These questions assess five key quality aspects: (Q1) Coherence, (Q2) Conversational quality, (Q3) Accuracy, (Q4) Comprehensiveness, and (Q5) Preference. 
We provide detailed guidelines and examples to the raters. 
In particular, we use Q3 to assess the reliability of generated outputs in avoiding hallucinations, as hallucinated content can undermine factual accuracy and user trust~\cite{maynez-2020-faithfulness}. 
Raters are instructed to label responses with a 0 if they detect potential hallucinations. 
The questions are listed in Appendix~\ref{app:appen_questions}, though specific details are omitted.

For generated queries, we asked human raters Q1 and Q2, along with an additional question: \textsl{Is the query relevant to the product listing?}.
Raters responded to this question with \texttt{Yes}, \texttt{No}, or \texttt{Somewhat Relevant}.

For enhanced listings, we assessed all five questions. In addition to human ratings, we also prompted the LLM model M5(Llama3.1-70b-instruct) to evaluate the enhanced listing data, aligning LLM values with human preferences in a demonstration setting. This is motivated by prior work highlighting the need for robust and scalable assessments of LLM alignment \cite{min-2024-exploring, sun-2024-principle, peng-2023-instruction, zheng-2024-judge}. Both human raters and the Llama model provided binary responses (\texttt{Yes} or \texttt{No} for Q1-Q4; for Q5, the response is \texttt{Yes} for enhanced listings or \texttt{No} for original listings).

\subsubsection{Do Llama Models Generate Diverse Synthetic Data?}
Training EBR models with diverse data improves generalization, enhancing performance on unseen data, such as new products and user queries~\cite{jagatap-2024-improve}. 
To assess data diversity, we used the Distinct-2 metric~\cite{li-2016-diversity}, which calculates the proportion of unique bigrams in generated text, helping to identify repetitive patterns and ensure varied, nuanced outputs for downstream retrieval and ranking tasks.
\begin{table}[!htb]
\centering
    \begin{tabular}{l r r r}
            \toprule
            Model & Queries & Listings \\ 
            \midrule
            M1 & 0.844 &0.716\\ 
            M2 & 0.923 &0.767\\ 
            M3 & 0.915 &0.784\\ 
            M4 & 0.910 &0.746\\
            M5 & 0.916 &0.756\\ 
            M6 & 0.889 &0.720\\ 
            M7 & 0.904 &0.746\\ 
            M8 & \textbf{0.957} & \textbf{0.824} & \\
        \bottomrule
        \end{tabular}
    \caption{Distinct-2 scores for randomly selected 200 samples of generated queries and enhanced listings.}
    \label{tab:dist-2}
    \Description{}
\end{table}
Table~\ref{tab:dist-2} presents Distinct-2 scores for queries and listings generated by models M1 through M8, highlighting lexical diversity across both synthetic data types. 
M8 stands out with the highest scores for both queries (0.957) and listings (0.824), suggesting it produces the most diverse content overall. 
M2 follows as the second-highest in query diversity (0.923), closely trailed by M5 and M3 with scores of 0.916 and 0.915, respectively. 
For listings, M3 achieves the second-highest diversity score (0.784), further demonstrating its effectiveness in varied output generation. 
In contrast, M1 displays the lowest Distinct-2 scores for both queries (0.844) and listings (0.716), indicating a more repetitive structure. Overall, M2 and M3 particularly excel in enhancing diversity, benefiting search and retrieval tasks by minimizing redundancy in generated content.

\begin{table}[!htb]
    \centering
    \small
    \begin{tabular}{l p{0.5cm} p{0.5cm} p{0.5cm} p{0.5cm} p{0.5cm} p{0.5cm} p{0.5cm}}
      \toprule
        & \multicolumn{2}{c}{Coherence} & \multicolumn{2}{c}{Conversational} & \multicolumn{3}{c}{Relevance} \\
        \cmidrule(lr){2-8}
        & Yes & No & Yes & No & Yes & No & SW. \\
        \midrule
        T1-basic & 97\% & 3\% & 90.4\% & 9.6\% & 59.9\% & 34\% & 6.1\% \\
        T2-detailed & 97.1\% & 2.9\% & 92.1\% & 7.9\% & 78.2\% & 5\% & 16.8\% \\
        \bottomrule
    \end{tabular}
    \caption{Human evaluation of generated queries using different templates. \textsl{SW.} represents \fqtperiod{somewhat relevant}.}
    \label{tab:human_templates}
\end{table}

\subsubsection{Do Prompts Affect Llama's Generation Capabilities?}

Table~\ref{tab:human_templates} shows the results for queries generated using two templates, T1-basic and T2-detailed.
For Coherence, both templates perform similarly, with nearly all queries rated as coherent (97\% for T1-basic, 97.1\% for T2-detailed). In terms of Conversational quality, T2-detailed slightly outperforms T1-basic (92.1\% vs. 90.4\%), suggesting it produces more natural, conversational queries.
The most notable difference appears in Relevance, where T2-detailed achieves a much higher relevance rating (78.2\%) than T1-basic (59.9\%), with fewer queries rated as irrelevant (5\% vs. 34\%) and a higher percentage as somewhat relevant (16.8\% vs. 6.1\%). This indicates that T2-detailed more effectively aligns queries with original product listings.
In summary, T2-detailed consistently outperforms T1-basic across all criteria, particularly in relevance, highlighting the effectiveness of detailed templates in generating high-quality, relevant synthetic queries.

\subsubsection{Do Synthetic Queries Retrieve Relevant Listings?}
To evaluate the relevance between generated queries and the given product listings, we deployed BM25-L and human evaluation.
Following \citet{sannigrahi-2024-synthetic}, we used the BM25-L retrieval model \cite{trotman-2015-bm}, with parameters \( k_1 = 1.5 \), \( b = 0.75 \), and \( \delta = 0.5 \), to assess how effectively each generated query retrieves its target listing. Since each query is designed to match a single relevant result, we measure retrieval performance using Reciprocal Rank (RR), which is particularly suited for cases with one relevant target. The RR is calculated as:
\[
\text{RR}(q, d) = \frac{1}{\text{rank}(q, d)}
\]
where \(\text{rank}(q, d)\) denotes the position of listing \(d\) for the generated query \(q\) under BM25-L. A higher RR indicates that the generated query is more effective in retrieving its target listing, reflecting the specificity and precision of the query.
To provide an overall measure of retrieval effectiveness, we reported the Mean Reciprocal Rank (MRR) by averaging the RR across the top-\(K\) (i.e. \(K\) = 5 in our case) generated queries per result. 

The MRR scores for models M1 through M8 are as follows: M1 (0.72), M2 (0.80), M3 (0.83), M4 (0.78), M5 (0.81), M6 (0.74), M7 (0.77), and M8 (0.86). 
Among these, M8 achieves the highest MRR of 0.86, suggesting superior query precision and alignment with intended targets, while M2 and M3 also perform well. 
These higher MRR values indicate that these models consistently retrieve the target result within the top ranks. In contrast, M1 and M6 exhibit lower MRR scores of 0.72 and 0.74, suggesting greater variability in query effectiveness and potential limitations in information relevance.

\begin{figure}[!htb]
    \centering
    \small
    \scalebox{0.85}{ 
    \begin{tikzpicture}
        \begin{axis}[
            width=6cm,
            height=5cm,
            ytick=data,
            yticklabels={M1, M2, M3, M4, M5, M6, M7, M8}, 
            xlabel={Percentage},
            xmin=0, xmax=100, 
            bar width=0.35cm,
            xbar stacked,
            nodes near coords,
            nodes near coords align={center},
            nodes near coords style={font=\color{black}\small, anchor=center}, 
            enlarge y limits=0.15,
            legend style={at={(0.5, -0.25)}, anchor=north, legend columns=-1, draw=none, font=\small},
            legend entries={Relevant, Somewhat Relevant, Not Relevant},
        ]
            \addplot+[xbar, fill=relevant, nodes near coords style={ color=white}, draw=none] 
                coordinates {(60.41,7) (77.67,6) (72.33,5) (72.11,4) (81.03,3) (66.20,2) (72.03,1) (75.33, 0)};
            \addplot+[xbar, fill=somewhat_relevant, draw=none] 
                coordinates {(29.29,7) (17.34,6) (21.40,5) (18.74,4) (16.52,3) (22.07,2) (22.61,1) (20.09, 0)};
            \addplot+[xbar, fill=not_relevant, draw=none, nodes near coords,
                nodes near coords style={font=\small, color=black, anchor=south, xshift=-3pt}]
                coordinates {(10.30,7) (4.99,6) (6.28,5) (9.15,4) (2.46,3) (11.73,2) (5.36,1) (4.59,0 )};
        \end{axis}
    \end{tikzpicture}
    } 
    \caption{Human evaluation for generated queries' relevance to the given product listings.}
    \label{fig:human_eval_q_rel}
    \Description{}
\end{figure}
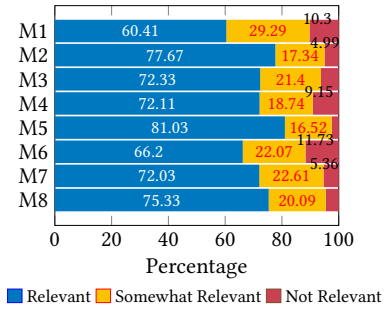

Figure~\ref{fig:human_eval_q_rel} presents human evaluations of relevance for generated queries across the eight models.
M5 achieves the highest relevance score at 81.03\%, followed by M2 and M8 at 77.67\% and 75.33\%, respectively, indicating that these models are particularly effective at generating queries closely aligned with product listings. 
Most models have low \fqt{Not Relevant} scores, though M6 and M1 exhibit slightly higher rates at 11.73\% and 10.30\%, indicating a need for further enhancement to improve query relevance.
The \fqt{Somewhat Relevant} category ranges from 16.52\% for M5 to 29.29\% for M1, with M7 also showing a relatively high score of 22.61\%, suggesting that its queries partially align with listings but lack full relevance. 
Upon deeper investigation, we found that \fqt{Somewhat Relevant} queries often relate to associative or tangentially relevant products. For instance, if the original listing describes a ``29.5" Solid Wood Dining Table'', a query such as ``small apartment affordable furniture'' is classified as \fqt{Somewhat Relevant} because it aligns with general categories of interest, indicating flexibility in how users search for related items--a positive sign for capturing broader user intent.

\begin{figure}[!htb]
    \centering
    \begin{subfigure}{0.48\linewidth}
        \centering
        \includegraphics[scale=0.75]{tikz/heatmap_human_eval_listings.tex}
        \caption{Human Evaluation}
    \end{subfigure}
    \hfill
    \begin{subfigure}{0.48\linewidth}
        \centering
        \includegraphics[scale=0.75]{tikz/heatmap_llm_eval_listings.tex}
        \caption{LLM Evaluation}
    \end{subfigure}
    \caption{Human and LLM evaluations for enhanced product listings.}
    \label{fig:performance-comparison}
    \Description{}
\end{figure}

\subsubsection{Do Human and LLM Evaluations Consistently Agree on Synthetic Data Quality?} 
\label{sec:do}

Results in Figure~\ref{fig:performance-comparison} present a strong overall alignment between human evaluations and LLM evaluations. Models generally perform well on Q1 to Q3, with most scores exceeding 80\%. 
However, M5 and M6 fall below 70\% on Q4 and Q5, indicating these questions pose greater challenges. 
M2, M3 and M8 demonstrate consistent performance across all criteria, with M8 slightly outperforming the other two in human evaluation. 
In contrast, M1, M5 and M6 score lower, especially on Q4 and Q5. 
High scores on Q3 indicate low hallucination rates, suggesting factual alignment with original product information.

We note that the LLM evaluation may introduce a bias toward LLM-generated data, potentially inflating alignment scores due to shared generation patterns. 
Among the evaluated models, M2 (Llama3-8b-instruct) demonstrates the most consistent and robust performance. 
This outcome may have partially been influenced by our study's timing, which was conducted before the release of Llama3.1. We extensively utilized Llama3 for prompt enhancement, which may have given it an advantage in this specific evaluation setup compared to the newer models.

\subsection{GenAI-Driven Data Augmentation for Enhancing EBR Models}

Based on the evaluation results, we select model M2 (Llama3-8b-instruct) for synthetic data augmentation in the EBR training pipeline. 
While M8 (Llama4.1-405b-instruct) outperforms M2 in some metrics, its lower inference Queries Per Second (QPS) limits its efficiency for large-scale synthetic data generation. 
M2 offers strong relevance and quality, particularly excelling in coherence and alignment with human assessments, and provides reliable high-quality data at a practical processing speed.

\subsubsection{Does Training Data Scale Affect Facebook Marketplace EBR Model Performance?}

Table~\ref{tab:res_ori} presents the performance metrics of EBR models trained with various combinations of original and synthetic data. The results indicate that synthetic data, particularly with the S3 strategy, boosts EBR performance across all metrics. S3 achieves the highest scores in every category with 100 million data points, reaching a PBC-o of 0.395, a PBC-r of 0.360, a RCR of 0.840, and a ROC\_AUC of 0.640. The S1 strategy also performs consistently well, especially in configurations with smaller data volumes, suggesting it may be more effective for moderate data sizes. For instance, at 100 million data points, S1 achieves a ROC\_AUC of 0.622, outperforming both S2 and the Original configurations. While S2 demonstrates competitive results, particularly in ROC\_AUC (0.637) at 100 million data points, its overall scores indicate slightly lower performance than S3. Across all strategies, increasing the data size from 50 million to 100 million yields incremental improvements, underscoring that scaling synthetic data enhances retrieval performance. The consistent improvement in PBC-o and ROC\_AUC across synthetic strategies further supports the advantage of synthetic augmentation over solely original data.

\begin{table}[!htb]
    \centering
    \small
    \begin{subtable}[t]{\linewidth}
        \centering
        \begin{tabular}{p{0.8cm} r r r r r}  
            \toprule
            Type & Data (m) & PBC-o & PBC-r & RCR & ROC\_AUC \\ 
            \midrule
            \rowcolor{decentgrey!60}
            Ori. & 50   & 0.340 & 0.302 & 0.750 & 0.580 \\
            \rowcolor{deepteal!60}
                 & 100  & 0.370 & 0.312 & 0.780 & 0.600 \\
            \midrule
            \rowcolor{decentgrey!60}
            S1       & 50   & 0.360 & 0.310 & 0.710 & 0.620 \\
            \rowcolor{deepteal!60}
                     & 100  & 0.390 & 0.316 & 0.760 & 0.622 \\
            \midrule
            \rowcolor{decentgrey!60}
            S2       & 50   & 0.355 & 0.310 & 0.720 & 0.625 \\
            \rowcolor{deepteal!60}
                     & 100  & 0.380 & 0.326 & 0.785 & 0.637 \\
            \midrule
            \rowcolor{decentgrey!60}
            S3       & 50   & 0.375 & 0.335 & 0.820 & 0.635 \\
            \rowcolor{deepteal!60}
                     & 100  & \textbf{0.395} & \textbf{0.360} & \textbf{0.840} & \textbf{0.640} \\
            \bottomrule
        \end{tabular}
        \caption{Performance of EBR models using various data types and sizes. Data scale is measured in millions.}
        \label{tab:res_ori}
    \end{subtable}
    
    \vspace{0.5cm} 
    
    \begin{subtable}[t]{\linewidth}
        \centering
        \begin{tabular}{p{0.7cm} p{0.7cm} l r r r r}  
            \toprule
            \multicolumn{2}{c}{Data} & \multirow{2}{*}{Strategy} & \multirow{2}{*}{PBC-o} & \multirow{2}{*}{PBC-r} & \multirow{2}{*}{RCR} & \multirow{2}{*}{ROC\_AUC} \\
            \cmidrule(lr){1-2}
            Ori. (m) & Syn. (m) & & & & & \\
            \midrule
            \rowcolor{decentgrey!60}
            25 & 25 & S1 & 0.355 & 0.315 & 0.760 & 0.615 \\
            \rowcolor{decentgrey!60}
               &    & S2 & 0.360 & 0.316 & 0.765 & 0.618 \\
            \rowcolor{decentgrey!60}
               &    & S3 & 0.365 & 0.320 & 0.770 & 0.622 \\
            \midrule
            25 & 50 & S1 & 0.375 & 0.328 & 0.790 & \textbf{0.630} \\
               &    & S2 & 0.380 & 0.326 & 0.780 & 0.628 \\
               &    & S3 & 0.385 & 0.332 & 0.785 & 0.635 \\
            \midrule
            25 & 100 & S1 & 0.400 & 0.335 & 0.810 & 0.645 \\
               &    & S2 & 0.405 & 0.338 & 0.815 & 0.648 \\
               &    & S3 & 0.410 & 0.342 & 0.825 & 0.650 \\
            \midrule
            50 & 25 & S1 & 0.365 & 0.318 & 0.780 & \textbf{0.622} \\
               &    & S2 & 0.370 & 0.315 & 0.775 & 0.620 \\
               &    & S3 & 0.375 & 0.312 & 0.785 & 0.619 \\
            \midrule
            \rowcolor{deepteal!60}
            50 & 50 & S1 & 0.385 & 0.328 & 0.800 & 0.635 \\
            \rowcolor{deepteal!60}
               &    & S2 & 0.390 & 0.326 & 0.795 & 0.633 \\
            \rowcolor{deepteal!60}
               &    & S3 & 0.395 & 0.332 & 0.805 & 0.638 \\
            \midrule
            50 & 100 & S1 & 0.405 & 0.340 & 0.830 & 0.655 \\
               &    & S2 & 0.410 & 0.342 & 0.835 & 0.658 \\
               &    & S3 & \textbf{0.415} & \textbf{0.345} & \textbf{0.850} & \textbf{0.660} \\
            \bottomrule
        \end{tabular}
        \caption{Performance metrics of EBR models trained with varying amounts of original (Ori.) and synthetic (Syn.) data, in millions.}
        \label{tab:res_ori_aug}
    \end{subtable}
    \caption{Performance metrics for EBR models using different data configurations. S1, S2, and S3 refer to synthetic data generation strategies introduced in Section~\ref{sec:strategies}. Data column indicates the total amount of data used to train EBR models. Bold values indicate the highest scores for each metric. The total data in highlighted rows under \textit{Ori.} and \textit{Syn.} in Table~\ref{tab:res_ori_aug} matches the \textit{Data} column in Table~\ref{tab:res_ori}.}
\end{table}

\subsubsection{Does Synthetic Data Enhance Facebook Marketplace EBR Model performance?}

In Table~\ref{tab:res_ori}, increasing the data amount from 50 million to 100 million across all strategies enhances model performance, showing a positive correlation between data volume and retrieval metrics. 
Table~\ref{tab:res_ori_aug} highlights the benefit of mixed original and synthetic data, where increased synthetic data enhances performance across metrics. With a total data size of 75 million (25 million original and 50 million synthetic), S3 achieves \( \text{PBC-o} = 0.385 \), \( \text{PBC-r} = 0.332 \), \( \text{RCR} = 0.785 \), and \( \text{ROC\_AUC} = 0.635 \), surpassing S1 and S2. 
At 125 million (50 original, 100 synthetic), S3 reaches peak values of \( \text{PBC-o} = 0.415 \), \( \text{PBC-r} = 0.345 \), \( \text{RCR} = 0.850 \), and \( \text{ROC\_AUC} = 0.660 \), underscoring its robustness in hybrid settings.
Moreover, in moderate data settings, S1 demonstrates competitive performance. 
In Table~\ref{tab:res_ori_aug}, with 75 million data points (25 million original and 50 million synthetic), S1 achieves a notable \( \text{ROC\_AUC} = 0.630 \), slightly outperforming S2 and closely trailing S3. 
At the 100-million data level (50 million original and 50 million synthetic), S1's scores remain strong, reaching \( \text{PBC-o} = 0.385 \) and \( \text{ROC\_AUC} = 0.635 \), suggesting S1 is particularly effective when synthetic data is moderate.

When the data amount is held constant, synthetic-only data consistently yields higher scores than mixed data, indicating that synthetic strategies S1, S2, and S3 provide more task-relevant information for EBR model training.
S3, in particular, achieves the highest scores across all metrics, suggesting its effectiveness in generating high-quality synthetic examples that capture nuanced contextual details beneficial for optimizing the model. In contrast, adding original data appears to introduce variability, slightly reducing focus on high-engagement and high-relevance attributes.

\section{Conclusion}

This research introduces a new framework, Aug2Search, for e-commerce platforms to leverage GenAI-generated synthetic data, enhancing EBR models and improving the alignment between user queries and product listings. 
Our approach demonstrates how synthetic data can drive higher-quality, diverse, and contextually relevant query-product matches, addressing key search challenges on platforms like Facebook Marketplace. Meta's GenAI models, particularly Llama3-8b, Llama3-70b, and Llama3.1-405b, have proven effective in generating synthetic queries and listings that enhance coherence, relevance, and diversity. 
Moreover, Aug2search demonstrates that increasing synthetic data volumes consistently improves EBR performance, with the S3 strategy (query generation after listing enhancement) yielding the best performance. 
These findings underscore the potential of GenAI-driven synthetic data to enhance large-scale e-commerce search, aligning closely with human standards for quality and relevance.

\bibliographystyle{ACM-Reference-Format}
\bibliography{Ruijie, Munindar, bach}

\appendix

\section{Questions}
\label{app:appen_questions}
\begin{enumerate}
    \item \textbf{Q1: Coherence}: Is the listing coherent, with no significant spelling errors? Minor grammatical issues that reflect the user's original input are acceptable, but major errors mark it as incoherent.
    \item \textbf{Q2: Conversational Quality}: Does the listing read naturally, as if written by a human?
    \item \textbf{Q3: Accuracy}: Does the revised listing accurately represent the original content without introducing false details (e.g., changing \fqt{6.1-inch screen} to \fqt{10-inch screen})?
    \item \textbf{Q4: Comprehensiveness}: Is the revised listing more comprehensive, including additional factual details without omitting key information from the original?
    \item \textbf{Q5: Preference}: Which listing (original or revised) is preferred?
\end{enumerate}

\section{Listing Enhancement Prompt}
\label{app:prompt_listings}

\begin{tcolorbox}[
    colback=gray!5!white,
    colframe=gray!75!black,
    title=Concise Product Description Generation Instructions,
    fonttitle=\bfseries,
    enhanced,
    breakable
]
Your task is to create a concise product description for a Facebook Marketplace listing. 
First, you should think step-by-step:
\begin{enumerate}
    \item Thoroughly analyze the \#\#\#Title and \#\#\#Description provided, correcting any grammar mistakes or spelling errors to ensure accurate understanding.
    \item Carefully identify the primary products and their attributes that accurately represent the \#\#\#Title and \#\#\#Description, focusing on the most relevant items to avoid confusion. 
\end{enumerate}

\smallskip
Then, follow these guidelines:
\begin{enumerate}
    \item Include all the features related to the product(s) provided in the given description.
    \item Avoid using superlatives and misleading adjectives that may lead to subjective interpretations.
    \item If the given description is not provided or is limited, generate features for the product(s) provided in the title to the best of your knowledge \textbf{WITHOUT} any additional information.
    \item If the given description contains some information, enrich it by sticking to relevant factual features about the product(s) \textbf{WITHOUT} any additional information.
    \item \textbf{NEVER} create placeholder text like ``insert features here''.
\end{enumerate}

Here is the listing:
\begin{itemize}
    \item \#\#\#Title: \{title\}
    \item \#\#\#Description: \{description\}
\end{itemize}

\end{tcolorbox}

\end{document}